# Quantum Confinement and Electronic Structure at the Surface of van der Waals Ferroelectric $\alpha$-In$_2$Se$_3$

Geoffroy Kremer[1,2], Aymen Mahmoudi[1], Adel M'Foukh[1], Meryem Bouaziz[1], Mehrdad Rahimi[3], Maria Luisa Della Rocca[3], Patrick Le Fèvre[4], Jean-Francois Dayen[5,6], François Bertran[4], Sylvia Matzen[1], Marco Pala[1], Julien Chaste[1], Fabrice Oehler[1], and Abdelkarim Ouerghi[1*]

[1]Université Paris-Saclay, CNRS, Centre de Nanosciences et de Nanotechnologies, 91120, Palaiseau, France

[2]Institut Jean Lamour, UMR 7198, CNRS-Université de Lorraine, Campus ARTEM, 2 allée André Guinier, BP 50840, 54011 Nancy, France

[3]Université Paris Cité, Laboratoire Matériaux et Phénomènes Quantiques, CNRS, UMR 7162, 75013 Paris, France.

[4]SOLEIL Synchrotron, L'Orme des Merisiers, Départementale 128, F-91190 Saint-Aubin, France

[5] Université de Strasbourg, IPCMS-CNRS UMR 7504, 23 Rue du Loess, 67034 Strasbourg, France
[6]Institut Universitaire de France, 1 rue Descartes, 75231 Paris cedex 05, France

## Abstract

Two-dimensional (2D) ferroelectric (FE) materials are promising compounds for next-generation nonvolatile memories, due to their low energy consumption and high endurance. Among them, $\alpha$-In$_2$Se$_3$ has drawn particular attention due to its in- and out-of-plane ferroelectricity, whose robustness has been demonstrated down to the monolayer limit. This is a relatively uncommon behavior since most bulk FE materials lose their ferroelectric character at the 2D limit due to depolarization field. Using angle resolved photoemission spectroscopy (ARPES), we unveil another unusual 2D phenomena appearing in 2H $\alpha$-In$_2$Se$_3$ single crystals, the occurrence of a highly metallic two-dimensional electron gas (2DEG) at the surface of vacuum-cleaved crystals. This 2DEG exhibits two confined states which correspond to an electron density of approximatively $10^{13}$ electrons/cm², also confirmed by thermoelectric measurements. Combination of ARPES and density functional theory (DFT) calculations reveals a direct band gap of energy equal to 1.3 +/- 0.1 eV, with the bottom of the conduction band localized at the center of the Brillouin zone, just below the Fermi level. Such strong n-type doping further supports the quantum confinement of electrons and the formation of the 2DEG.

**Corresponding Author:** abdelkarim.ouerghi@c2n.upsaclay.fr

Two-dimensional (2D) van der Waals (vdW) III−VI semiconductors have drawn intense attention due to their interesting electronic properties[1]. Among these materials, α-In$_2$Se$_3$ in its hexagonal (2H) and rhombohedral (3R) polytypes[2] shows a great potential for a wide variety of applications in electronics, photonics and even thermoelectricity[3]. These two polytypes of α-In$_2$Se$_3$ combine exotic ferroelectricity and a rather remarkable band structure[4]. Moreover, α-In$_2$Se$_3$ represents an intriguing platform when coupled with other 2D layers, allowing to tune electronic, magnetic and optical properties, through band structure engineering[5,6].

The interest of layered III-VI semiconductors from a fundamental and technological point of view, relies on unveiling the electronic properties of their various crystalline phases. In particular, the understanding of their electronic band structure, isolated[2] or in contact with other materials, is crucial for future device engineering[7]. In the literature, layered III-VI semiconductors, such as In$_2$Se$_3$, Ga$_2$Se$_3$, In$_2$Te$_3$ and others, are mainly studied for their ability to retain a stable ferroelectric (FE) phase at room temperature, even down to the monolayer[8]. This unusual and robust FE character should allow for the integration of FE functions in other 2D materials, by vdW heterostructuring. This rich physics requires intensive investigations, as the band structure and electronic properties of In$_2$Se$_3$ depend on the layer thickness[9,10], on the particular atomic arrangement between individual layers (polytype) and on atoms arrangement inside each layer (crystal phase)[2]. Despite the large interest in layered 2D III-VI semiconductors, many details of their optical and electronic properties are still unknown. In particular, there is no experimental report on the electronic band structure of 2H and 3R phases of α-In$_2$Se$_3$, even at room temperature.

In this work, we explore in detail the electronic band structure of 2H α-In$_2$Se$_3$ single crystals by angle resolved photoemission spectroscopy (ARPES), revealing the presence of a direct band gap of approximatively 1.3 eV. We further unveil the occurrence of a two-dimensional electron gas (2DEG) at the surface of FE α-In$_2$Se$_3$, at the center of the Brillouin zone (BZ), close to the Fermi level ($E_F$), due to an intrinsic strong *n*-type doping. 2DEGs are usually found at the surface of oxides[11–13] or in carefully designed quantum wells and interfaces in planar III-V or II-VI semiconductors[14,15]. Due to their very high electron mobilities[16], 2DEGs are ideal platforms for exploring fundamental physics or high performance electronics. Here, a 2DEG combines with the natural FE character of α-In$_2$Se$_3$, which represents an intriguing merging of physical properties[8]. The identification of a 2DEG in III-VI 2D semiconductors represents a key step that can be exploited to improve electronic, photoconductive and thermoelectric functionalities of a wide range of nano-devices. Any advance in this domain could open routes towards FE 2DEGs whose transport properties could be electrostatically switched in a non-volatile way[17] by means of external stimuli such as an electric field[18,19] or ultrafast light pulses[20]. This outstanding property, together with its relatively large electronic band gap, makes α-In$_2$Se$_3$ even exploitable for applications in devices operating at room temperature.

## Results and Discussion

### Structural and stacking properties.

Among III–VI semiconductors, In$_2$Se$_3$ can crystallize into various structures, combining several phases (α, β, γ, δ, and κ)[2,7,21] and polytypes. The α-In$_2$Se$_3$ phase studied here is a standard layered vdW structure with well-defined covalently-bonded monolayers in the plane, weakly coupled to each other in the out-of-plane direction by vdW forces. Each monolayer consists of two indium atoms with different (covalent) coordination polyhedra: In(1) is coordinated tetrahedrally by four Se atoms while In(2) is octahedrally coordinated by six Se (figure 1(a)). The inner structure of the individual monolayer thus decomposes into five sub-layers, in a Se-In$^1$-Se-In$^2$-Se sequence along the vertical direction. Figure 1a shows a schematic of three repetitions of such a quintuple sub-layer (QL) unit stacked vertically, in which the thickness of individual QL is about ~1.2 nm$^2$. In addition, α-In$_2$Se$_3$ may show several polytypes depending on the relative position and in-plane orientation of each QL in the full unit cell. The polytype 2H α-In$_2$Se$_3$ shows hexagonal symmetry with space group P6$_3$mc and, as shown in figure 1(a), corresponds to the standard AB configuration, in which the A- and B-type QL are related by a 180° in plane rotation. As visible in figure 1(a), the first and third QL (odd number) have identical in-plane atomic coordinates. Figure 1(b) shows a schematic of the bulk and surface-projected hexagonal BZ of 2H α-In$_2$Se$_3$, where the two high-symmetry planes (ALH) and (ΓMK) of the three-dimensional (3D) BZ are shown in green and red, respectively.

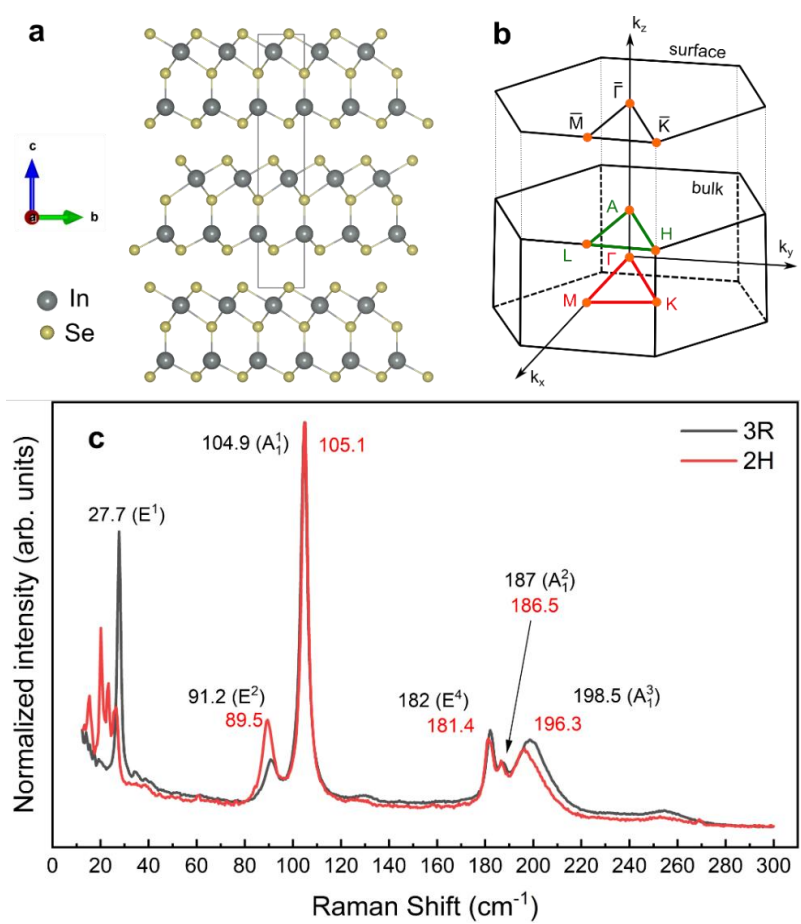

**Figure 1:** (a) Side view of the surface crystal structure (the first three quintuple layers (QLs)) of 2H α-In$_2$Se$_3$ plotted with the VESTA visualization program. The black line shows the conventional unit cell. (b) Corresponding three-dimensional (3D) Brillouin zone (BZ) and its 2D projection in the (001) plane. (c) High resolution micro-Raman spectroscopy (HR-RS) spectra for 2H (red) and 3R (black) stacked α-In$_2$Se$_3$ crystals using 532 nm laser.

As already demonstrated, micro-Raman spectroscopy can be used to identify the fingerprint of α-In$_2$Se$_3$ crystals phase and polytype[2]. Here, room-temperature micro-Raman spectra of commercial 2H and 3R α-In$_2$Se$_3$ crystals

(HQ graphene) were obtained by using a Horiba Labram micro Raman microscope with 532 nm laser excitation[22,23] for the comparison. As shown in Figure 1(c), multiple Raman peaks are observed near 25, 90, 105, 182, 187, 197 and 255 cm$^{-1}$ in both 2H (red curve) and 3R (black curve) α-In$_2$Se$_3$ crystals, which can be assigned to the E and A modes of α-In$_2$Se$_3$[2]. We observe that the E$^2$ mode of 2H α-In$_2$Se$_3$ is more prominent than that of 3R. The intensity ratio between E and A modes can thus be used as a criterion for identifying the 3R and 2H α-In$_2$Se$_3$ polytypes. We also observe similar dependence of the mode ratios as a function of the excitation energy (see Supplementary material Figure S1).

**Dispersive electronic states.**

We now turn our attention to the electronic properties of 2H α-In$_2$Se$_3$. Figure 2 presents ARPES measurements performed at T = 80 K using different incident photon energy, as an efficient tool to explore the full 3D electronic band structure of a material[24]. In Figure 2 (a), we show a ($k_x$, hv) ARPES dependence for a given binding energy (BE), namely BE=5 eV below $E_F$, where we observe three bands named B$_1$, B$_2$ and B$_3$. Whereas B$_1$ and B$_2$ show a non-dispersive behavior as a function of the photon energy, the B$_3$ band has a dispersive nature evidenced by the maximum of intensity at hv = 68 and 87 eV, and the lack of spectral weight at hv = 77 eV. Such variation with the photon energy is indicative of dispersion along the out-of-plane direction in the 3D BZ. It is worthwhile to note that these three photon energies (68, 77 and 87 eV) correspond to high symmetry points along the vertical direction, which we assign to A, Γ and A points. This is confirmed by the analysis of Figure 2 (b-d), where we plot the (E, $k_x$) ARPES maps for each of these three photon energies. By comparing these ARPES measurements with density functional theory (DFT) calculations (see Supplementary Material Figure S2), we can unambiguously confirm the nature of each high symmetry directions, namely H-A-H (green plane in Figure 1(b)) and K-Γ-K (red plane in Figure 1(b)) using hv = 68/87 eV and hv = 77 eV, respectively. Using an inner potential of $V_0$=10 eV, we estimate that the out-of-plane periodicity along $k_z$ (distance between two A points) equals approximatively 0.55 Å$^{-1}$.

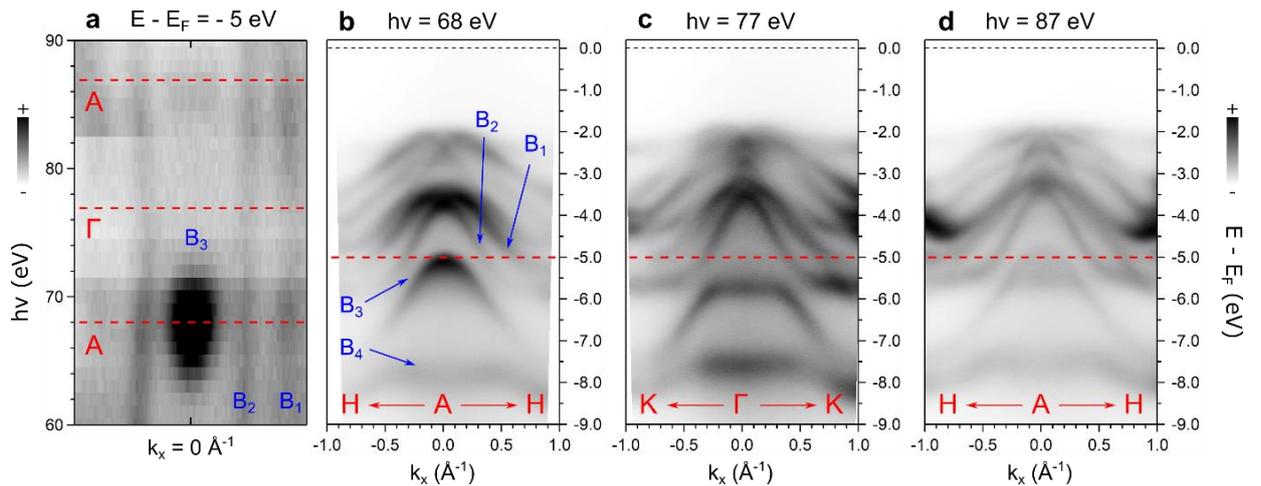

**Figure 2:** (a) Photon energy dependent ARPES dispersion for 2H α-In$_2$Se$_3$ at $k_y$ = 0 Å$^{-1}$ and E - $E_F$ = - 5 eV. The red dashed lines exhibit the photon energies corresponding to the A and Γ points of the 3D BZ. Corresponding ARPES spectra along the H-A-H and K-Γ-K high symmetry directions (green and red planes in Fig.1(b)) for (b) hv = 68 eV, (c) hv = 77 eV and (d) hv = 87 eV.

In details, when comparing Figure 2 (b) and (c), we can observe that the $B_3$ band shows a strong $k_z$ dispersion, absent for $B_2$ and $B_1$ bands. By comparing with DFT, the $B_3$ band shows a maximum at the A point and a minimum at the Γ point which is the most pertinent criterion to identify our high symmetry direction. We can notice that a similar behavior occurs for the $B_4$ band at higher BE. DFT calculations (Figure S2) show that the $B_3$ and $B_4$ bands have mainly a $p_z$ (out-of-plane) character near Γ whereas $B_1$ and $B_2$ bands have a $p_x$ and $p_y$ (in-plane) character. This symmetry feature is in line with the experimental dispersions along the out-of-plane direction of the BZ (Figure 2 (a)). Going away from the center of the BZ, $B_3$ and $B_4$ bands recover an in-plane character and a reduced $k_z$ dispersion. These bands symmetries are further confirmed by their polarization dependence response, probed by polarized ARPES experiments presented in Supplementary Material (Figure S3). In particular, the suppression of photoemission spectral weight in a linear vertical polarization configuration at normal emission is typical of $p_z$ orbitals[25,26].

We now focus on ARPES measurements in the (ALH) plane of the 3D BZ (Figure 1(b)). In Figure 3(a) we present the superposition of isoenergetic contours in the ($k_x$, $k_y$) plane obtained between BE=0 eV ($E_F$) and BE=6 eV. Each individual isoenergetic contour is reported separately in Figure 3(b-g). All the isoenergetics contours clearly show the hexagonal symmetry of the BZ, as expected from the crystal cell symmetry in the direct space. In Figure 3(g), we have superimposed the projected 2D hexagonal BZ on the isoenergetic contour, showing an excellent agreement using the reference in plane lattice parameter of the crystal[2] (a= 4.0 Å). At this BE (6 eV below $E_F$), we can resolve some spectral weight at the H point, no photoemission intensity at the L point and a circular structure around the A point, which disperses as a hole band when probing at even lower BE. In Figure 3(h) we plot the full ARPES dispersion (sum of linear vertical and linear horizontal polarizations) along the A-H-L high symmetry direction of the BZ. We compare the experimental results by superimposing the DFT calculation of the 2H phase in Figure 3(i). For the lower part of the band structure, the agreement is excellent, with the exception of the $B_3$ band showing some $k_z$ broadening near the Γ point at the used photon energy, hν=87 eV (see also corresponding calculations in Figure S2).

More interestingly, we experimentally resolve a lack of photoemission intensity between approximatively 1.7 eV BE and $E_F$ which is well described by DFT at BE=1.7eV but it is not matching the expected DFT band structure calculation near BE=0 eV. We ascribe this lack of intensity to the electronic band gap of the material. Therefore, the photoemission intensities below 1.7 eV corresponds to the valence band (VB) of the material and the intensities in the vicinity of $E_F$, visible as a single spot at the A point at the topmost isoenergetic map in Figure 3(a) and on ARPES maps in Figure 3(b), correspond to the signature of the conduction band (CB). The position of $E_F$ inside the CB observed experimentally can be explained by a strong electron doping of the crystal, on which we focus in the next section. We do not observe experimentally any other signature of the CB at any other points of the BZ, which is expected to disperse at higher energy according to DFT calculations. In our case, those unoccupied states are simply non-accessible using static ARPES measurements.

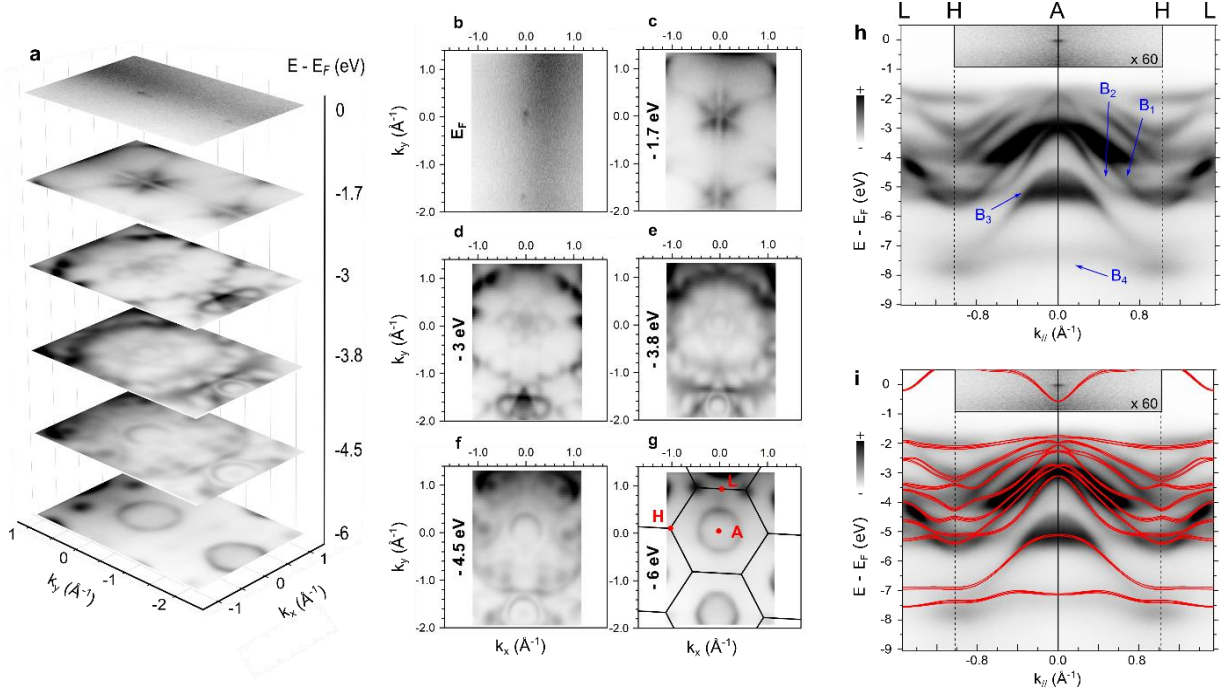

**Figure 3:** (a) Constant ARPES energy surfaces for 2H α-In$_2$Se$_3$ (hν = 85 eV) with 2D projections taken at E - E$_F$ equals (b) 0 eV (c) - 1.7 eV (d) - 3 eV (e) - 3.8 eV (f) - 4.5 eV and (g) - 6 eV. In panel (g), black lines correspond to the 2D hexagonal BZ obtained using the in-plane cell parameter of α-In$_2$Se$_3$ (a = 4.0 Å). (h) High-resolution ARPES spectrum along the L-H-A-H-L high symmetry direction obtained by summing the LH and LV polarizations showing the dispersions of both the valence and the conduction bands. (i) Same as panel (h) with superimposed DFT calculations obtained in the GGA approximation (a rigid energy shift has been applied to fit the ARPES dispersion).

## Determination of the electronic band gap.

Since the most striking features of the electronic band structure of α-In$_2$Se$_3$ are localized at the center of the BZ, we now discuss ARPES spectra acquired at lower photon energy in order to increase the energy and momentum resolutions. These additional ARPES measurements performed at hν = 20 eV are also corresponding to the investigation of the (AHL) plane since the difference of k$_z$ between this photon energy and hν = 87 eV is an integer multiple of the above determined out-of-plane periodicity. The raw and normalized spectra are shown in Figure 4 (a) and (b), respectively. On the same normalized spectrum, we can now directly resolve the dispersions of both the VB and the CB, leading to the conclusion that the minimum of the CB and the maximum of the VB are localized at the center of the BZ, in agreement with above described DFT calculations. It is further evidenced in Figure 4 (c) which corresponds to a zoom in of panel (b) (red dashed rectangle). Consequently, we can directly evaluate a higher approximation of the material band gap by plotting the density of state at the center of the BZ, as shown in Figure 4 (d). We conclude that 2H α-In$_2$Se$_3$ exhibits a direct band gap of approximatively E$_F$ = 1.3 +/- 0.1 eV by considering both the VB and CB leading edges, in excellent agreement with DFT calculations using the GW approximation[27] or photoluminescence measurements[7].

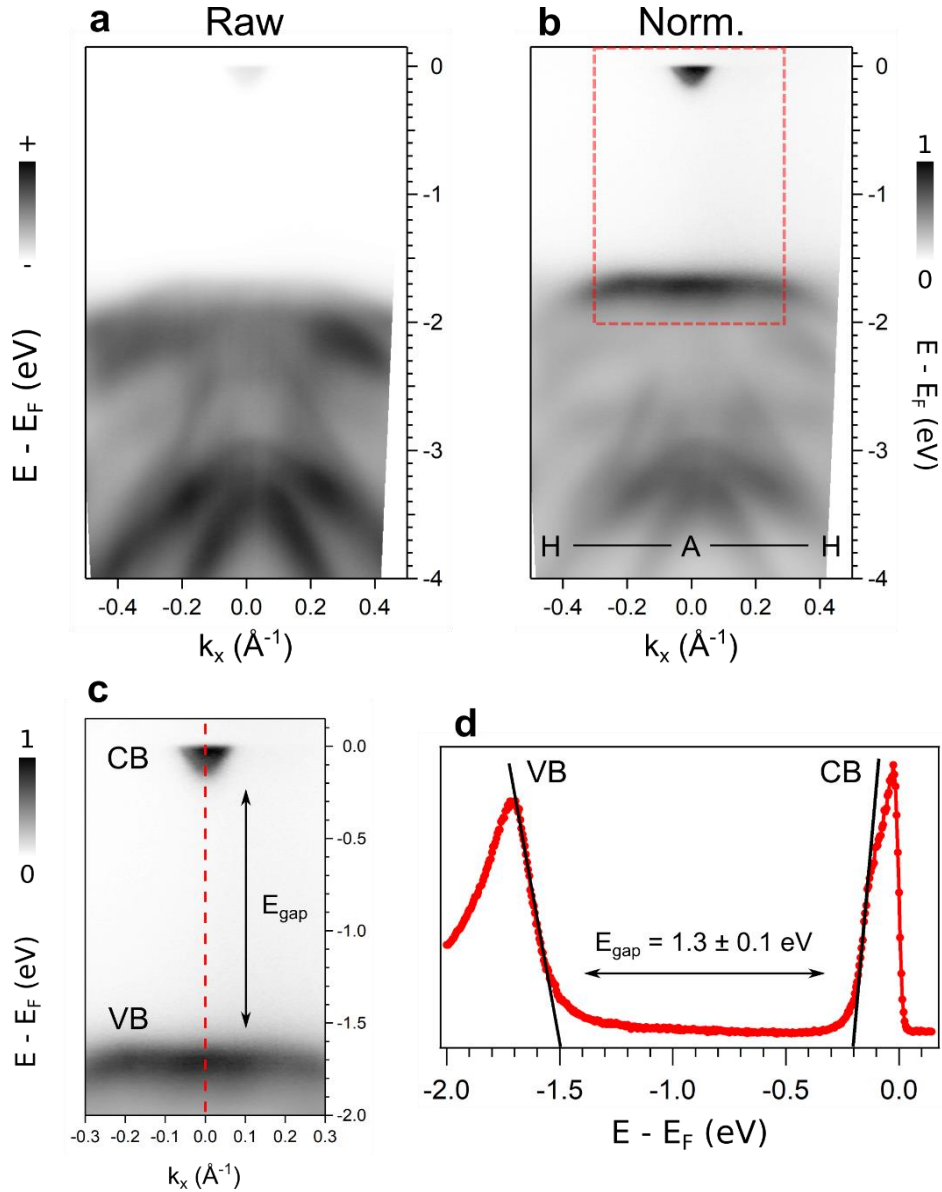

**Figure 4:** ARPES spectrum of 2H α-In$_2$Se$_3$ (hν = 20 eV) along the H-A-H high symmetry direction. (a) Raw and (b) normalized data. The red dashed rectangle in panel (b) corresponds to a zoom-in which is represented in panel (c) and shows evidence of the direct band gap of the material which is estimated by plotting the normal emission energy distribution curve (red line) in panel (d).

## Quantum confinement and emergence of a 2D electron gas.

A more detailed inspection of the CB dispersion is done in Figure 5(a). It shows that it does not consist of a single band but rather it involves at least two bands. This feature is typical of surface charge accumulation layers, which is a key-signature of quantum confinement and it is interpreted as the fingerprint of the occurrence of a 2D electron gas (2DEG) detected by ARPES measurements[28,29]. While 2DEGs are usually created by engineering band offsets and charge transfer at the heterointerfaces of (planar) III-V or II-VI semiconductors, it is known that high charge density can also be induced at bare surfaces of various transition metal oxides or group III nitrides (InN)[29]. Such surface 2DEGs are particularly well-suited to direct investigations by surface sensitive ARPES experiments. We can quantify the dispersive nature of these specific bands using a parabolic dispersion. We obtain an effective mass (m*) of 0.1 $m_0$ for both bands, with their energy minimum localized at 60 and 140 meV BE, respectively (red and blue dashed lines in Figure 5). With parabolic dispersion and low effective masses, these two bands match the signature of confined quantum well states in the CB. The 2DEG is particularly visible at 20 eV due to a combination of favorable cross section and surface sensitivity at this energy. It is important to mention that, as expected for a 2D state, this 2DEG does not disperse as a function of the photon energies we explored, as it is evidenced in Figure S4 for hν = 12 / 20 / 31 eV. These photon energies correspond respectively to an off high symmetry direction (12 eV), H-A-H direction (20 eV) and K-Γ-K line (31 eV). If the observed state was the "true" CB as calculated by bulk DFT (see Figure S5), a strong modulation of the position of the band should be observed: it is not the case according to our ARPES measurements and consequently goes in the direction of the evidence of a 2DEG. The formation of a quantum-confined state at the surface is a common phenomenon in semiconductor physics, which relates to the highly doped character of the material[15,30]. In our case, the 2H α-$In_2Se_3$ crystal has a strong n-type character, since $E_F$ is located far above the top of the VB[30]. At the interface with the vacuum, we have a downward band bending of the CB, leading to the formation of quantum confined states at the surface of the sample. We have computed the potential well and the corresponding accumulation charge at an $In_2Se_3$ surface by numerically solving the self-consistent Poisson-Schrödinger equation (figure (5(b)). The parameters of the simulation are $T$=300 K, m*=0.1$m_0$, $\varepsilon_{sc}$=17$\varepsilon_0$. A donor sheet density of 6x$10^{13}$ cm$^{-2}$ was assumed to be smeared into a thin layer at the surface (0 < Z < 3nm). As shown in the figure, the lowest two confined energy levels were found at energies $E_1$=-0.185eV and $E_2$=-0.055eV with respect to the Fermi level $E_F$ = 0. The shape of the absolute value of the corresponding wavefunctions is also shown to illustrate their spatial extension. These surface states share the dispersive features of conventional 2DEG, as shown in Figure 5(b) are compatible with ARPES measurements. Therefore, our measurements unveil the formation of a 2DEG at the surface of the layered FE 2H α-$In_2Se_3$ semiconductor. For the sake of clarity, it is important to note that the value of the band gap extracted above is consequently a higher estimation since it corresponds to the energy difference between the top of the VB and the bottom of the lowest quantum well state, and not exactly to the difference between the top of the VB and the bottom of the CB. Nevertheless, this is a good approximation since the band bending in our case is likely equal to a few hundreds of meV.

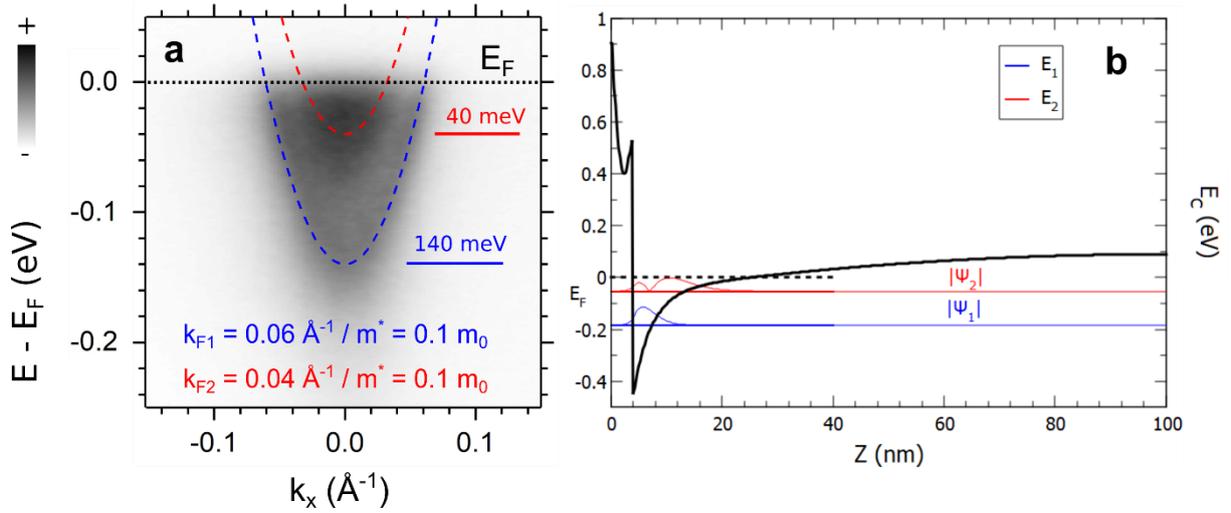

**Figure 5:** (a) High-resolution ARPES spectrum of 2H α-In$_2$Se$_3$ (hν = 20 eV) obtained in the vicinity of the Fermi energy showing two parabolic bands. Red and blue dashed lines correspond to a fit with a nearly free electron dispersion. (b) Corresponding Surface potential (black line) and wave functions (red and blue lines) extracted using the QW model and experimental energy parameters obtained from (a).

We can finally estimate the carrier density of the formed 2DEG from the area enclosed by each Fermi surface. Here, we compute the sum of the two individual 2D Luttinger volume $n_{2D} = k_F^2/2\pi$ of the bands (respectively $5.7 \times 10^{12}$ and $2.5 \times 10^{12}$ electrons/cm²) with $k_F$ being the Fermi wave vector of the two sub-bands[31]. We consequently obtain a total electron density of approximatively $10^{13}$ electrons/cm². To strengthen our analysis, we have additionally fabricated 2H α-In$_2$Se$_3$ based devices for standard field-effect and thermoelectric measurements, with flakes thicknesses in the range 85-115 nm (see Supplementary material Figure S6). The flakes inserted in the devices have been exfoliated from single crystals analogous to the ones used for ARPES measurements. Electronic transport, measured at T=35°C under high-vacuum (P~$10^{-7}$ mbar), reveals non-linear current-voltage characteristics with no clear gate dependence, indicating charge injection far from the bandgap region in agreement with the Fermi level position extracted by ARPES. Most importantly, we measure a constant negative thermopower in the whole explored gate voltage range (equal to 90–220μV/K in absolute value depending on the sample), indicating electrons as majority charge carriers. By using the Mott formula and the extracted effective mass, we find a density of charge carrier $n_{2D}$ in the range of $10^{13}$ electrons/cm² for the different measured devices, in agreement with the previous estimation.

Such a high value of the electron density is compatible with the electrostatic origin of the band bending. The exact magnitude of the surface band bending is difficult to extract from our ARPES experiments since the acquired dispersions represent an average of the energy shift over several unit cells along the depth probed by photoemission. Furthermore, understanding the origin of the corresponding electric field in 2H α-In$_2$Se$_3$ 2DEG is an important step towards achieving carrier density control. While it is well established that 2DEGs form due to potential gradient arising from work function mismatch in conventional semiconductors, the surface origin of both the attractive confining potential and the excess charge carriers in FE 2H α-In$_2$Se$_3$ sample is interesting for the future works. As one of the most promising 2D materials, α-In2Se3 has a thickness-dependent bandgap varying from monolayer to bulk. As such we may expect some of the properties of the 2DEG at the surface of α-In2Se3 to also be thickness-dependent. This is not the case of 2DEG found in transition metal oxides and group III nitrides. Further measurements on the 3R α-In$_2$Se$_3$ phase (see preliminary data in Figure S7) are also desired, in particular a comprehensive 2H/3R comparison to evaluate the effects of staking order on the electronic band structure.

## Conclusions and Perspectives

In summary, we have used ARPES to resolve the electronic structure of FE 2H α-In$_2$Se$_3$ single crystals, confirming its semiconducting nature with a direct band gap of 1.3 +/- 0.1 eV, with the bottom of the CB localized just below the Fermi level at the center of the BZ. We interpret this observation as a pronounced n-type doping that we ascribe to hypothetic surface defects that could be resolved in future works by means of local probe spectroscopy such as scanning tunneling microscopy. Such high doping promotes the formation of an electron accumulation layer at the surface, with quantum confinement signature in the CB. The formed quantum well hosts a 2DEG with quantized energy levels ($E_1$ and $E_2$) that we directly observed thanks to high resolution ARPES. This 2DEG is robust under synchrotron illumination and time and is associated to an electron density of approximatively $10^{13}$ electrons/cm². Further investigations such as surface alkali doping, systematic $k_z$ dispersion and exploration of the homogeneity of the 2DEG with sub-micron spatial resolution could be interesting for its better understanding. It will also be investigated both by optical spectroscopy and electrical measurements in future studies. Our work consequently reveals that the α-In$_2$Se$_3$ surface hosts an interesting combination of ferroelectricity and confined electronic states (2DEG), opening the route for further studies to precisely control this emerging 2DEG and to understand how it is intertwined with the intrinsic α-In$_2$Se$_3$ ferroelectric nature[32].

## Materials and Methods

**Experiments.** ARPES experiments were performed at the CASSIOPEE beamline of the SOLEIL synchrotron light source. The CASSIOPEE beamline is equipped with a Scienta R4000 hemispherical electron analyzer and the incident photon beam was focused into a <100 µm spot on the sample surface. High-quality samples from the "HQ Graphene" company were cleaved at room temperature at a base pressure better than $5 \times 10^{-9}$ mbar. All the photoemission data were extracted from these cleaved surfaces without any annealing. The experiment was performed at T = 80 K with an energy resolution better than 15 meV. The µ-Raman measurements were conducted at room temperature, using a commercial confocal Horiba micro-Raman microscope with a ×100 objective and a 532 / 633 nm laser excitation. The laser beam was focused onto a small spot having a diameter of ~1 µm on the sample.

**Computational details.** The theoretical calculations were carried out by using first principles calculations based on density functional theory (DFT) by means of the Quantum ESPRESSO suite[33]. We used a fully relativistic pseudopotential and non-collinear simulations to consider the spin-orbit interaction and adopted norm conserving pseudopotentials based on the Perdew-Burke-Ernzerhof (PBE)[34] exchange-correlation functional. Van der Waals interactions were considered in the calculations with non-local van der Waals functional vdW-DF3[35]. The self-consistent solution was obtained by adopting a 10x10x10 Monkhorst-Pack grid and a cutoff energy of 80 Ry. Cell parameters and atomic positions were relaxed according to a convergence threshold for forces and energy of $10^{-3}$ and $10^{-8}$ (a.u.), respectively.

## Associated Content

The Supporting Information is available free of charge on the …. at DOI:.

Readers can find additional Raman spectroscopy measurements on both 2H and 3R α-In$_2$Se$_3$ single crystals, complementary DFT calculations and polarization dependent band structure measurements, ARPES measurements of the 2DEG on 2H α-In$_2$Se$_3$ at different photon energies, preliminary ARPES measurements on 3R α-In$_2$Se$_3$, and electric/thermoelectric measurements on a 2H α-In$_2$Se$_3$ based device.

**DATA AVAILABILITY**: The datasets generated during and/or analyzed during the current study are available from the corresponding author on reasonable request.

**ACKNOWLEDGMENTS:** We acknowledge the financial support by MagicValley (ANR-18-CE24-0007), Graskop (ANR-19-CE09-0026), 2D-on-Demand (ANR-20-CE09-0026), and MixDferro (ANR-21-CE09-0029) grants, as well as the French technological network RENATECH.

**Competing interests:** The authors declare no competing interests.

## REFERENCES

(1) Geim, a K.; Grigorieva, I. V. Van Der Waals Heterostructures. *Nature* **2013**, *499* (7459), 419–425. https://doi.org/10.1038/nature12385.

(2) Liu, L.; Dong, J.; Huang, J.; Nie, A.; Zhai, K.; Xiang, J.; Wang, B.; Wen, F.; Mu, C.; Zhao, Z.; Gong, Y.; Tian, Y.; Liu, Z. Atomically Resolving Polymorphs and Crystal Structures of In2Se3. *Chem. Mater.* **2019**, *31* (24), 10143–10149. https://doi.org/10.1021/acs.chemmater.9b03499.

(3) Huang, Y.-T.; Chen, N.-K.; Li, Z.-Z.; Wang, X.-P.; Sun, H.-B.; Zhang, S.; Li, X.-B. Two-Dimensional In2Se3: A Rising Advanced Material for Ferroelectric Data Storage. *InfoMat* **2022**, *4* (8), e12341. https://doi.org/10.1002/inf2.12341.

(4) Shi, W.; Yu, S.; Liu, P.; Fan, W.; Luo, H.; Song, S. Near-Infrared Photoluminescent Flowerlike α-In2Se3 Nanostructures from a Solvothermal Treatment. *Chem. Eng. J.* **2013**, *225*, 474–480. https://doi.org/10.1016/j.cej.2013.03.066.

(5) Ayadi, T.; Debbichi, L.; Badawi, M.; Said, M.; Kim, H.; Rocca, D.; Lebègue, S. An Ab Initio Study of the Ferroelectric In2Se3/Graphene Heterostructure. *Phys. E Low-Dimens. Syst. Nanostructures* **2019**, *114*, 113582. https://doi.org/10.1016/j.physe.2019.113582.

(6) Ayadi, T.; Debbichi, L.; Badawi, M.; Said, M.; Rocca, D.; Lebègue, S. An Ab Initio Study of the Electronic Properties of the Ferroelectric Heterostructure In2Se3/Bi2Se3. *Appl. Surf. Sci.* **2021**, *538*, 148066. https://doi.org/10.1016/j.apsusc.2020.148066.

(7) Balakrishnan, N.; Steer, E. D.; Smith, E. F.; Kudrynskyi, Z. R.; Kovalyuk, Z. D.; Eaves, L.; Patanè, A.; Beton, P. H. Epitaxial Growth of γ-InSe and α, β, and γ-In2Se3 on ε-GaSe. *2D Mater.* **2018**, *5* (3), 035026. https://doi.org/10.1088/2053-1583/aac479.

(8) Xue, F.; Hu, W.; Lee, K.-C.; Lu, L.-S.; Zhang, J.; Tang, H.-L.; Han, A.; Hsu, W.-T.; Tu, S.; Chang, W.-H.; Lien, C.-H.; He, J.-H.; Zhang, Z.; Li, L.-J.; Zhang, X. Room-Temperature Ferroelectricity in Hexagonally Layered α-In2Se3 Nanoflakes down to the Monolayer Limit. *Adv. Funct. Mater.* **2018**, *28* (50), 1803738. https://doi.org/10.1002/adfm.201803738.

(9) Lv, B.; Yan, Z.; Xue, W.; Yang, R.; Li, J.; Ci, W.; Pang, R.; Zhou, P.; Liu, G.; Liu, Z.; Zhu, W.; Xu, X. Layer-Dependent Ferroelectricity in 2H-Stacked Few-Layer α-In2Se3. *Mater. Horiz.* **2021**, *8* (5), 1472–1480. https://doi.org/10.1039/D0MH01863E.

(10) Wan, S.; Li, Y.; Li, W.; Mao, X.; Zhu, W.; Zeng, H. Room-Temperature Ferroelectricity and a Switchable Diode Effect in Two-Dimensional α-In2Se3 Thin Layers. *Nanoscale* **2018**, *10* (31), 14885–14892. https://doi.org/10.1039/C8NR04422H.

(11) Ohtomo, A.; Hwang, H. Y. A High-Mobility Electron Gas at the LaAlO3/SrTiO3 Heterointerface. *Nature* **2004**, *427* (6973), 423–426. https://doi.org/10.1038/nature02308.

(12) Reyren, N.; Thiel, S.; Caviglia, A. D.; Kourkoutis, L. F.; Hammerl, G.; Richter, C.; Schneider, C. W.; Kopp, T.; Rüetschi, A.-S.; Jaccard, D.; Gabay, M.; Muller, D. A.; Triscone, J.-M.; Mannhart, J. Superconducting Interfaces Between Insulating Oxides. *Science* **2007**, *317* (5842), 1196–1199. https://doi.org/10.1126/science.1146006.

(13) Caviglia, A. D.; Gabay, M.; Gariglio, S.; Reyren, N.; Cancellieri, C.; Triscone, J.-M. Tunable Rashba Spin-Orbit Interaction at Oxide Interfaces. *Phys. Rev. Lett.* **2010**, *104* (12), 126803. https://doi.org/10.1103/PhysRevLett.104.126803.

(14) Bahramy, M. S.; King, P. D. C.; de la Torre, A.; Chang, J.; Shi, M.; Patthey, L.; Balakrishnan, G.; Hofmann, P.; Arita, R.; Nagaosa, N.; Baumberger, F. Emergent Quantum Confinement at Topological Insulator Surfaces. *Nat. Commun.* **2012**, *3* (1), 1159. https://doi.org/10.1038/ncomms2162.

(15) Bianchi, M.; Guan, D.; Bao, S.; Mi, J.; Iversen, B. B.; King, P. D. C.; Hofmann, P. Coexistence of the Topological State and a Two-Dimensional Electron Gas on the Surface of Bi2Se3. *Nat. Commun.* **2010**, *1* (1), 128. https://doi.org/10.1038/ncomms1131.

(16) Fukutani, K.; Sato, T.; Galiy, P. V.; Sugawara, K.; Takahashi, T. Tunable Two-Dimensional Electron Gas at the Surface of Thermoelectric Material In4Se3. *Phys. Rev. B* **2016**, *93* (20), 205156. https://doi.org/10.1103/PhysRevB.93.205156.

(17) Zheng, D.; Zhang, J.; He, X.; Wen, Y.; Li, P.; Wang, Y.; Ma, Y.; Bai, H.; N. Alshareef, H.; Zhang, X.-X. Electrically and Optically Erasable Non-Volatile Two-Dimensional Electron Gas Memory. *Nanoscale* **2022**, *14* (34), 12339–12346. https://doi.org/10.1039/D2NR01582J.

(18) Varotto, S.; Nessi, L.; Cecchi, S.; Sławińska, J.; Noël, P.; Petrò, S.; Fagiani, F.; Novati, A.; Cantoni, M.; Petti, D.; Albisetti, E.; Costa, M.; Calarco, R.; Buongiorno Nardelli, M.; Bibes, M.; Picozzi, S.; Attané, J.-P.; Vila, L.; Bertacco, R.; Rinaldi, C. Room-Temperature Ferroelectric Switching of Spin-to-Charge Conversion in Germanium Telluride. *Nat. Electron.* **2021**, *4* (10), 740–747. https://doi.org/10.1038/s41928-021-00653-2.

(19) Si, M.; Saha, A. K.; Gao, S.; Qiu, G.; Qin, J.; Duan, Y.; Jian, J.; Niu, C.; Wang, H.; Wu, W.; Gupta, S. K.; Ye, P. D. A Ferroelectric Semiconductor Field-Effect Transistor. *Nat. Electron.* **2019**, *2* (12), 580–586. https://doi.org/10.1038/s41928-019-0338-7.

(20) Kremer, G.; Maklar, J.; Nicolaï, L.; Nicholson, C. W.; Yue, C.; Silva, C.; Werner, P.; Dil, J. H.; Krempaský, J.; Springholz, G.; Ernstorfer, R.; Minár, J.; Rettig, L.; Monney, C. Field-Induced Ultrafast Modulation of Rashba Coupling at Room Temperature in Ferroelectric α-GeTe(111). *Nat. Commun.* **2022**, *13* (1), 6396. https://doi.org/10.1038/s41467-022-33978-3.

(21) Zhang, F.; Wang, Z.; Liu, L.; Nie, A.; Gong, Y.; Zhu, W.; Tao, C. Atomic-Scale Visualization of Polar Domain Boundaries in Ferroelectric In2Se3 at the Monolayer Limit. *J. Phys. Chem. Lett.* **2021**, *12* (49), 11902–11909. https://doi.org/10.1021/acs.jpclett.1c03251.

(22) Ernandes, C.; Khalil, L.; Almabrouk, H.; Pierucci, D.; Zheng, B.; Avila, J.; Dudin, P.; Chaste, J.; Oehler, F.; Pala, M.; Bisti, F.; Brulé, T.; Lhuillier, E.; Pan, A.; Ouerghi, A. Indirect to Direct Band Gap Crossover in Two-Dimensional WS2(1−x)Se2x Alloys. *Npj 2D Mater. Appl.* **2021**, *5* (1), 1–7. https://doi.org/10.1038/s41699-020-00187-9.

(23) Shahmanesh, A.; Romanin, D.; Dabard, C.; Chee, S.-S.; Gréboval, C.; Methivier, C.; Silly, M. G.; Chaste, J.; Bugnet, M.; Pierucci, D.; Ouerghi, A.; Calandra, M.; Lhuillier, E.; Mahler, B. 2D Monolayer of the 1T' Phase of Alloyed WSSe from Colloidal Synthesis. *J. Phys. Chem. C* **2021**, *125* (20), 11058–11065. https://doi.org/10.1021/acs.jpcc.1c02275.

(24) Henck, H.; Pierucci, D.; Zribi, J.; Bisti, F.; Papalazarou, E.; Girard, J.-C.; Chaste, J.; Bertran, F.; Le Fèvre, P.; Sirotti, F.; Perfetti, L.; Giorgetti, C.; Shukla, A.; Rault, J. E.; Ouerghi, A. Evidence of Direct Electronic Band Gap in Two-Dimensional van Der Waals Indium Selenide Crystals. *Phys. Rev. Mater.* **2019**, *3* (3), 034004. https://doi.org/10.1103/PhysRevMaterials.3.034004.

(25) Kremer, G.; Alvarez Quiceno, J. C.; Lisi, S.; Pierron, T.; González, C.; Sicot, M.; Kierren, B.; Malterre, D.; Rault, J. E.; Le Fèvre, P.; Bertran, F.; Dappe, Y. J.; Coraux, J.; Pochet, P.; Fagot-Revurat, Y. Electronic Band Structure of Ultimately Thin Silicon Oxide on Ru(0001). *ACS Nano* **2019**, *13* (4), 4720–4730. https://doi.org/10.1021/acsnano.9b01028.

(26) Yukawa, R.; Yamamoto, S.; Ozawa, K.; D'Angelo, M.; Ogawa, M.; Silly, M. G.; Sirotti, F.; Matsuda, I. Electronic Structure of the Hydrogen-Adsorbed SrTiO3(001) Surface Studied by Polarization-Dependent Photoemission Spectroscopy. *Phys. Rev. B* **2013**, *87* (11), 115314. https://doi.org/10.1103/PhysRevB.87.115314.

(27) Debbichi, L.; Eriksson, O.; Lebègue, S. Two-Dimensional Indium Selenides Compounds: An Ab Initio Study. *J. Phys. Chem. Lett.* **2015**, *6* (15), 3098–3103. https://doi.org/10.1021/acs.jpclett.5b01356.

(28) Olszowska, N.; Lis, J.; Ciochon, P.; Walczak, Ł.; Michel, E. G.; Kolodziej, J. J. Effect of a Skin-Deep Surface Zone on the Formation of a Two-Dimensional Electron Gas at a Semiconductor Surface. *Phys. Rev. B* **2016**, *94* (11), 115305. https://doi.org/10.1103/PhysRevB.94.115305.

(29) Jovic, V.; Moser, S.; Ulstrup, S.; Goodacre, D.; Dimakis, E.; Koch, R.; Katsoukis, G.; Moreschini, L.; Mo, S.-K.; Jozwiak, C.; Bostwick, A.; Rotenberg, E.; Moustakas, T. D.; Smith, K. E. How Indium Nitride Senses Water. *Nano Lett.* **2017**, *17* (12), 7339–7344. https://doi.org/10.1021/acs.nanolett.7b02985.

(30) Zhang, Z.; Yates, J. T. Jr. Band Bending in Semiconductors: Chemical and Physical Consequences at Surfaces and Interfaces. *Chem. Rev.* **2012**, *112* (10), 5520–5551. https://doi.org/10.1021/cr3000626.

(31) Walker, S. M.; Bruno, F. Y.; Wang, Z.; de la Torre, A.; Riccó, S.; Tamai, A.; Kim, T. K.; Hoesch, M.; Shi, M.; Bahramy, M. S.; King, P. D. C.; Baumberger, F. Carrier-Density Control of the SrTiO3 (001) Surface 2D Electron Gas Studied by ARPES. *Adv. Mater.* **2015**, *27* (26), 3894–3899. https://doi.org/10.1002/adma.201501556.

(32) Bréhin, J.; Trier, F.; Vicente-Arche, L. M.; Hemme, P.; Noël, P.; Cosset-Chéneau, M.; Attané, J.-P.; Vila, L.; Sander, A.; Gallais, Y.; Sacuto, A.; Dkhil, B.; Garcia, V.; Fusil, S.; Barthélémy, A.; Cazayous, M.; Bibes, M. Switchable Two-Dimensional Electron Gas Based on Ferroelectric Ca:SrTiO3. *Phys. Rev. Mater.* **2020**, *4* (4), 041002. https://doi.org/10.1103/PhysRevMaterials.4.041002.

(33) Giannozzi, P.; Baroni, S.; Bonini, N.; Calandra, M.; Car, R.; Cavazzoni, C.; Ceresoli, D.; Chiarotti, G. L.; Cococcioni, M.; Dabo, I.; Corso, A. D.; Gironcoli, S. de; Fabris, S.; Fratesi, G.; Gebauer, R.; Gerstmann, U.; Gougoussis, C.; Kokalj, A.; Lazzeri, M.; Martin-Samos, L.; Marzari, N.; Mauri, F.; Mazzarello, R.; Paolini, S.; Pasquarello, A.; Paulatto, L.; Sbraccia, C.; Scandolo, S.; Sclauzero, G.; Seitsonen, A. P.; Smogunov, A.; Umari, P.; Wentzcovitch, R. M. QUANTUM ESPRESSO: A Modular and Open-Source Software Project for Quantum Simulations of Materials. *J. Phys. Condens. Matter* **2009**, *21* (39), 395502. https://doi.org/10.1088/0953-8984/21/39/395502.

(34) Perdew, J. P.; Burke, K.; Ernzerhof, M. Generalized Gradient Approximation Made Simple. *Phys. Rev. Lett.* **1996**, *77* (18), 3865–3868. https://doi.org/10.1103/PhysRevLett.77.3865.

(35) Chakraborty, D.; Berland, K.; Thonhauser, T. Next-Generation Nonlocal van Der Waals Density Functional. *J. Chem. Theory Comput.* **2020**, *16* (9), 5893–5911. https://doi.org/10.1021/acs.jctc.0c00471.

# Supplementary Material for: Quantum Confinement and Electronic Structure at the Surface of van der Waals Ferroelectric α-In$_2$Se$_3$

Geoffroy Kremer[1,2], Aymen Mahmoudi[1], Adel M'Foukh[1], Meryem Bouaziz[1], Mehrdad Rahimi[3], Maria Luisa Della Rocca[3], Patrick Le Fèvre[4], Jean-Francois Dayen[5,6], François Bertran[4], Sylvia Matzen[1], Marco Pala[1], Julien Chaste[1], Fabrice Oehler[1], and Abdelkarim Ouerghi[1*]

[1]Université Paris-Saclay, CNRS, Centre de Nanosciences et de Nanotechnologies, 91120, Palaiseau, France

[2]Institut Jean Lamour, UMR 7198, CNRS-Université de Lorraine, Campus ARTEM, 2 allée André Guinier, BP 50840, 54011 Nancy, France

[3]Université Paris Cité, Laboratoire Matériaux et Phénomènes Quantiques, CNRS, UMR 7162, 75013 Paris, France.

[4]SOLEIL Synchrotron, L'Orme des Merisiers, Départementale 128, F-91190 Saint-Aubin, France

[5] Université de Strasbourg, IPCMS-CNRS UMR 7504, 23 Rue du Loess, 67034 Strasbourg, France

[6]Institut Universitaire de France, 1 rue Descartes, 75231 Paris cedex 05, France

**Corresponding Author:** abdelkarim.ouerghi@c2n.upsaclay.fr

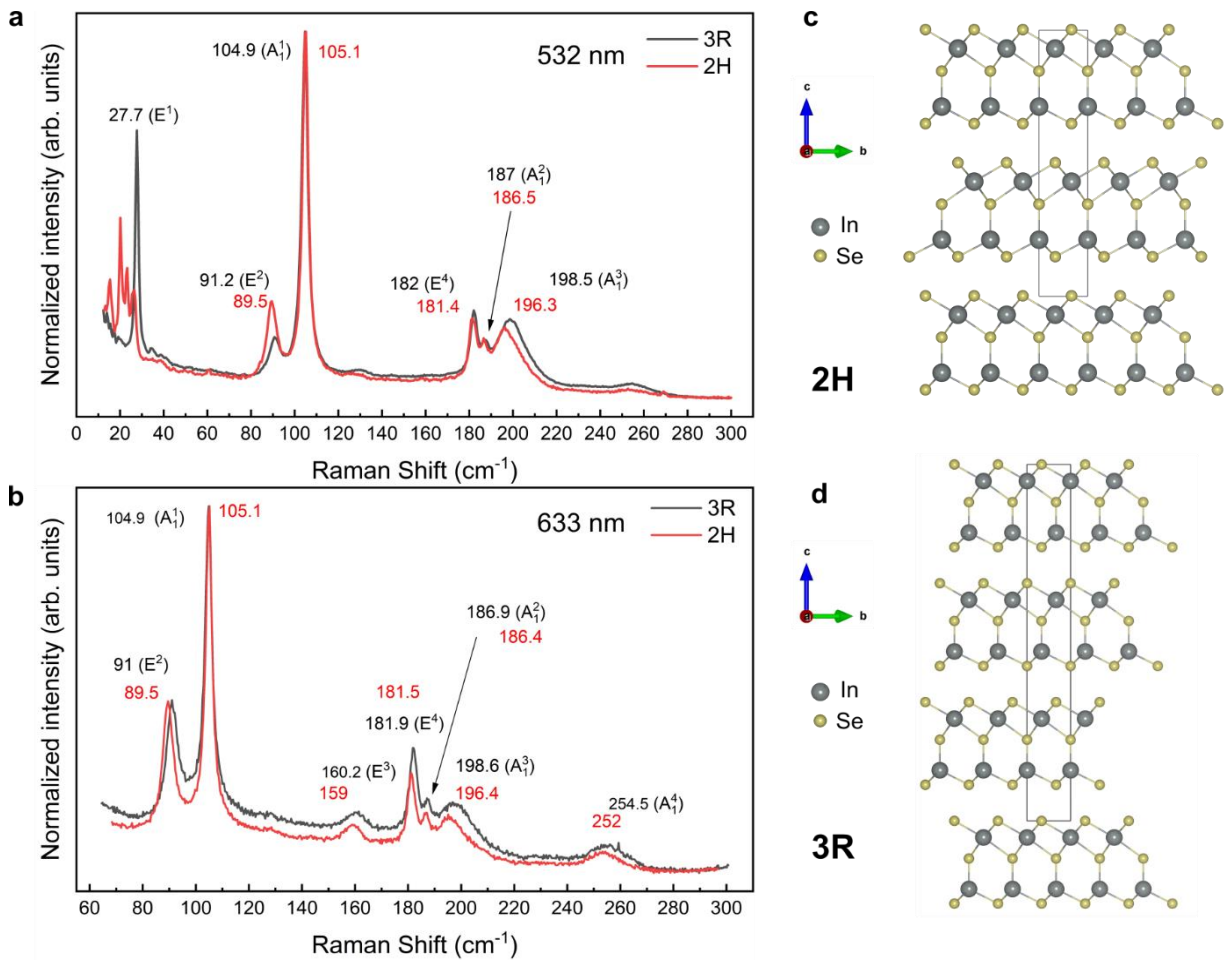

**Figure S1**: (a,b) Room-temperature high resolution Raman spectroscopy (HR-RS) spectra for 2H (red) and 3R (black) stacked α-In$_2$Se$_3$ crystals using 532 nm and 633 nm excitations, respectively. (c,d) Side view of the surface crystal structure of 2H and 3R α-In$_2$Se$_3$. The black line shows the conventional unit cell.

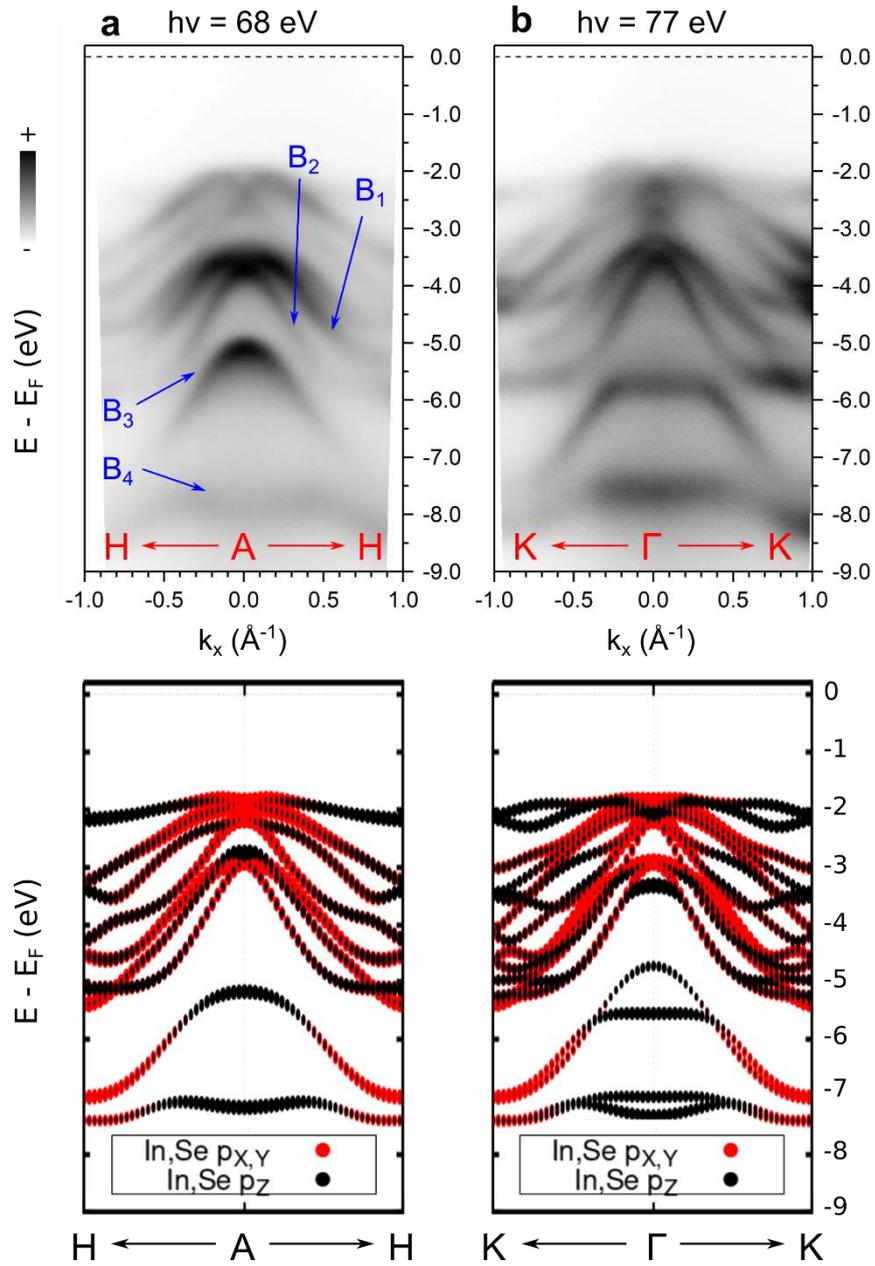

**Figure S2**: (a) ARPES spectra of 2H α-In$_2$Se$_3$ (hν = 68 eV) along the H-A-H high symmetry direction (top) and corresponding DFT calculation showing the bands symmetries (p$_{x,y}$ : red color and p$_z$ : black color). Only the VB contribution is shown in the calculations for the sake of clarity. (b) Same for the K-Γ-K high symmetry direction (hν = 77 eV). A rigid energy shift of the DFT calculations has been applied to correspond to ARPES dispersions.

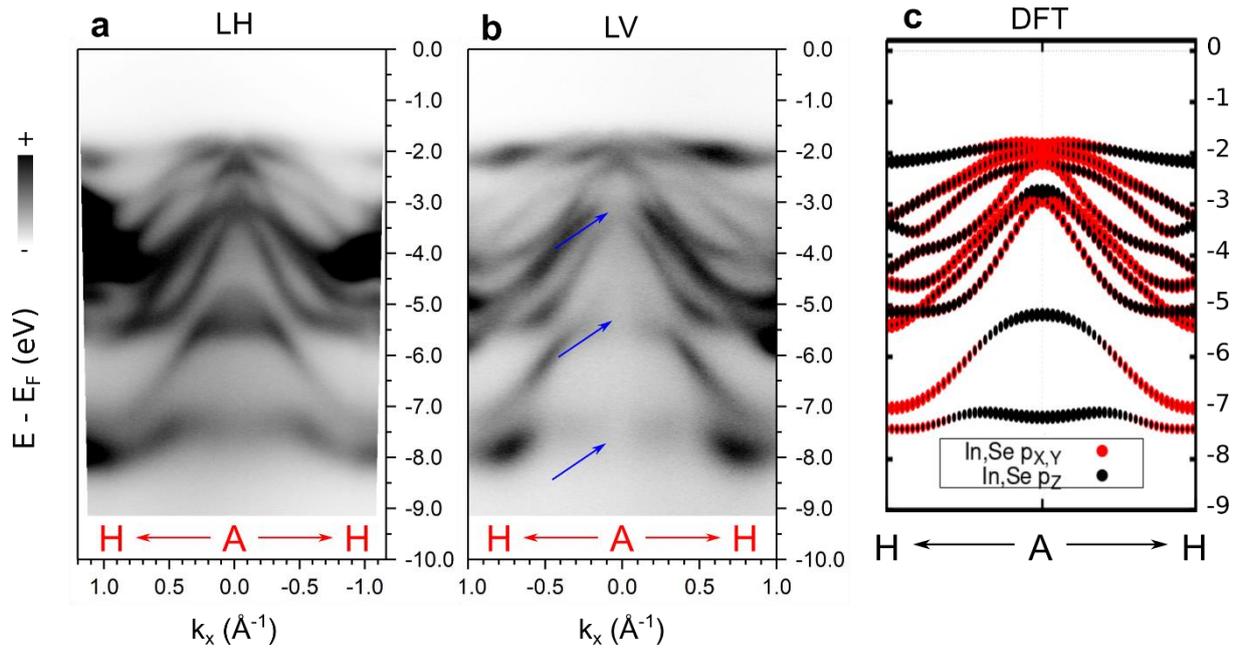

**Figure S3**: ARPES spectra of 2H α-$In_2Se_3$ (hν = 85 eV) along the H-A-H high symmetry direction for (a) linear horizontal (LH) and (b) linear vertical (LV) polarizations. The blue arrows highlight the suppression of photoemission intensity at normal emission due to matrix elements effect of the photoemission process. (c) Corresponding DFT calculations showing the orbitals symmetries of the bands, in particular the $p_z$ character of the bands where the suppression of intensity is observed (only the VB contribution is shown in the calculations for the sake of clarity). A rigid energy shift of the DFT calculations has been applied to correspond to ARPES dispersions.

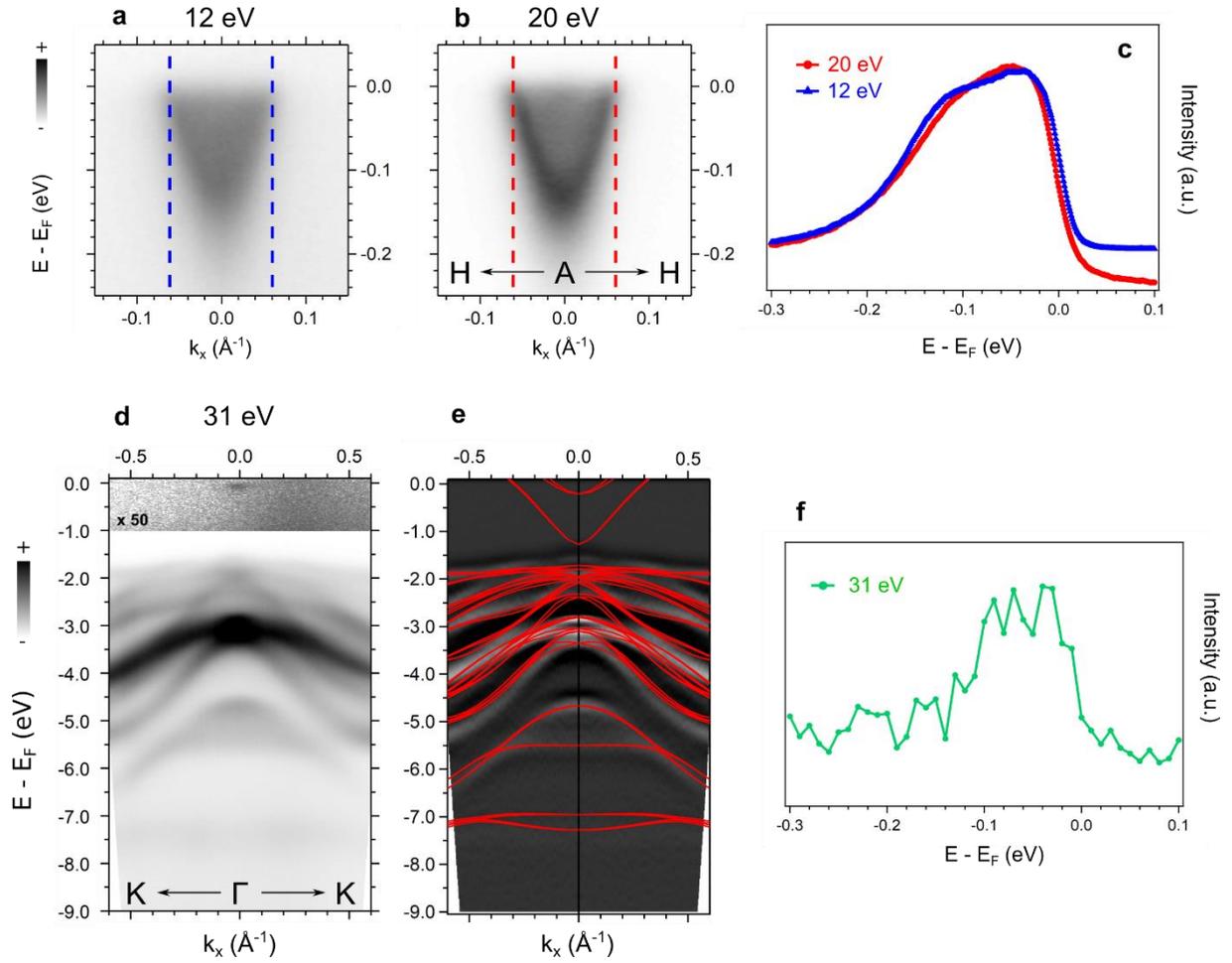

**Figure S4**: ARPES spectra of 2H α-In$_2$Se$_3$ closed to the Fermi level for (a) hν = 12 eV and (b) hν = 20 eV. Note that these spectra have been measured at a different position compared to data shown in Figure 4 of the main text. (c) Corresponding energy distribution curves (EDCs) around the normal emission showing the non-dispersing character of the 2DEG in this photon energy range. (d) ARPES spectrum of 2H α-In$_2$Se$_3$ (hν = 31 eV) along the K-Γ-K high symmetry direction. (e) Corresponding second derivative spectrum of panel (d) with superimposed DFT calculations obtained in the GGA approximation (a rigid energy shift has been applied to fit the ARPES dispersion). Note that the band gap is underestimated in this approximation compared to experiments and more advanced GW calculations. (f) EDC in the vicinity of the Fermi level and taken around the normal emission evidencing again the non-dispersing 2DEG as a function of the photon energy.

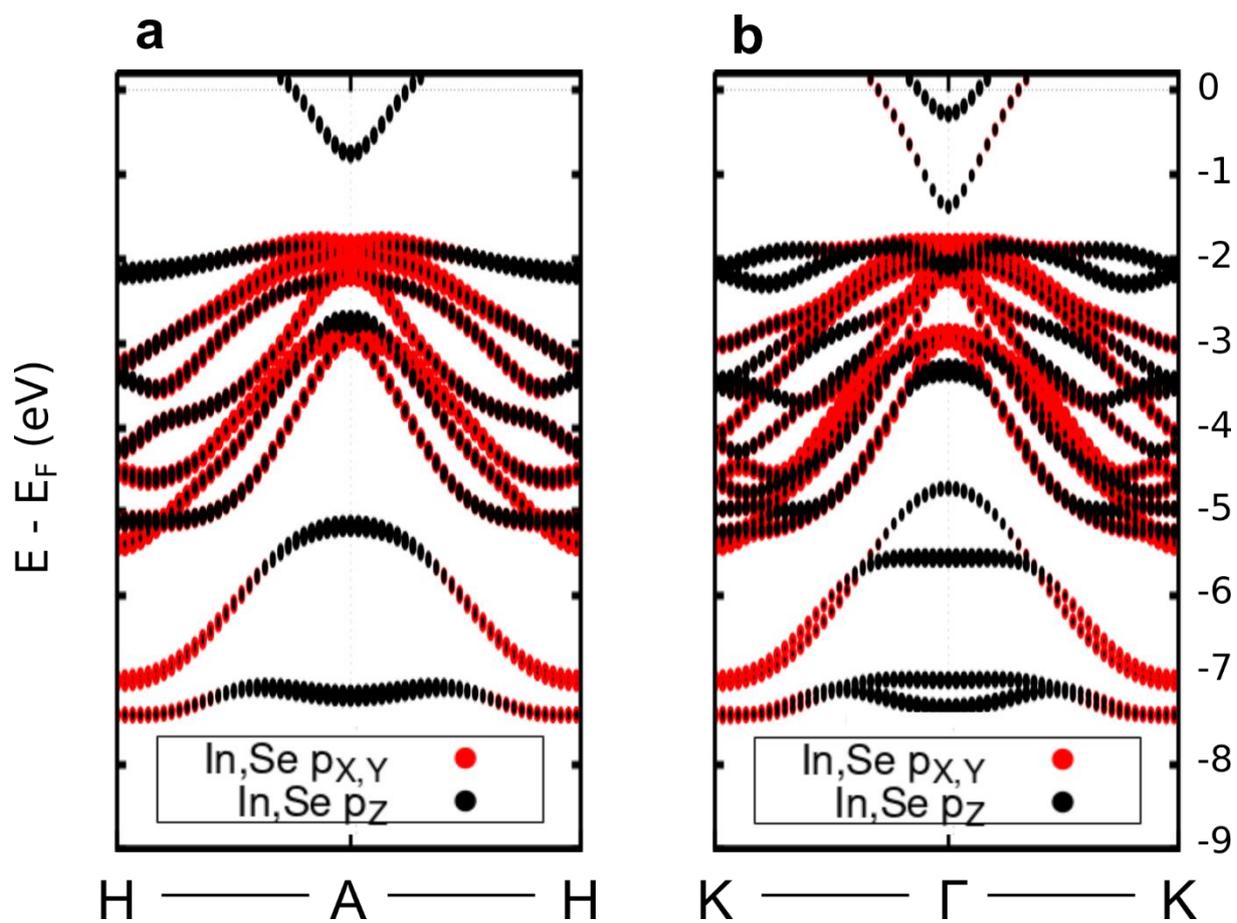

**Figure S5**: DFT calculations showing the bands dispersions and symmetries ($p_{x,y}$ : red color and $p_z$ : black color) for both the CB and the VB along the (a) H-A-H and (b) K-Γ-K high symmetry directions.

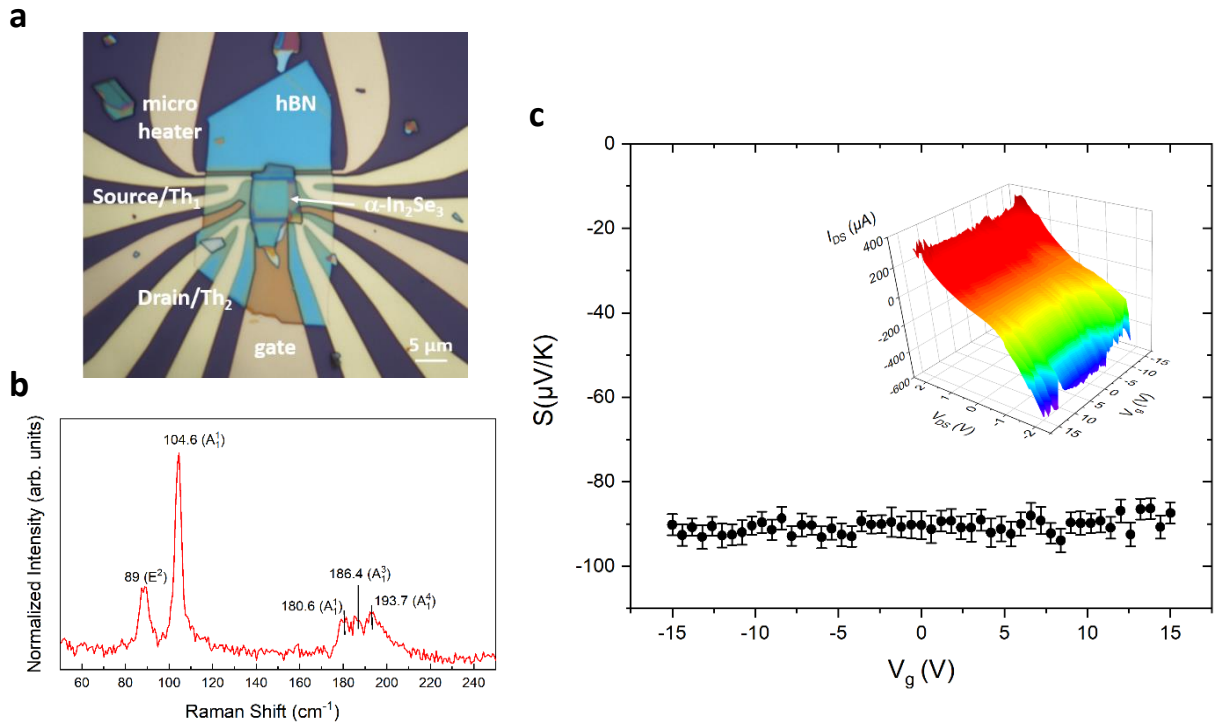

**Figure S6**: (a) 2H α-In$_2$Se$_3$ based device for standard DC electric and thermoelectric measurements. A local gate electrode is nanofabricated over a Si/SiO$_2$ substrate close to a micro-heater. A 50-nm thick hBN flake is used as dielectric spacer between the gate and the top most 110-nm thick 2H α-In$_2$Se$_3$ flake. Both hBN and In$_2$Se$_3$ have been exfoliated from bulk single crystals and successively transferred. Two metallic nanowires are nanofabricated between the hBN and 2H α-In$_2$Se$_3$ transfer steps, 6 μm-far from each other. Each nanowire is connected to four leads and used as local source/drain electrode and local thermometer (Th$_1$, Th$_2$). The adopted device geometry, where the transfer of the 2H α-In$_2$Se$_3$ flake occurs as the last step of the fabrication process, assures a reduced exposition to clean room environment of the freshly prepared surface. (b) Raman spectra (Witec microscope with 532 nm laser excitation) of the 2H α-In$_2$Se$_3$ flake of the device in (a) showing the characteristics peaks of the In$_2$Se$_3$ α-phase, with a prominent E$^2$ mode. (c) Measured thermopower, $S$, as a function of the gate voltage, V$_g$, showing a constant negative value, indicative of electron doping. Inset: 3D plot of the IV characteristics of the 2H α-In$_2$Se$_3$ based device as a function of the gate voltage and the source/drain voltage.

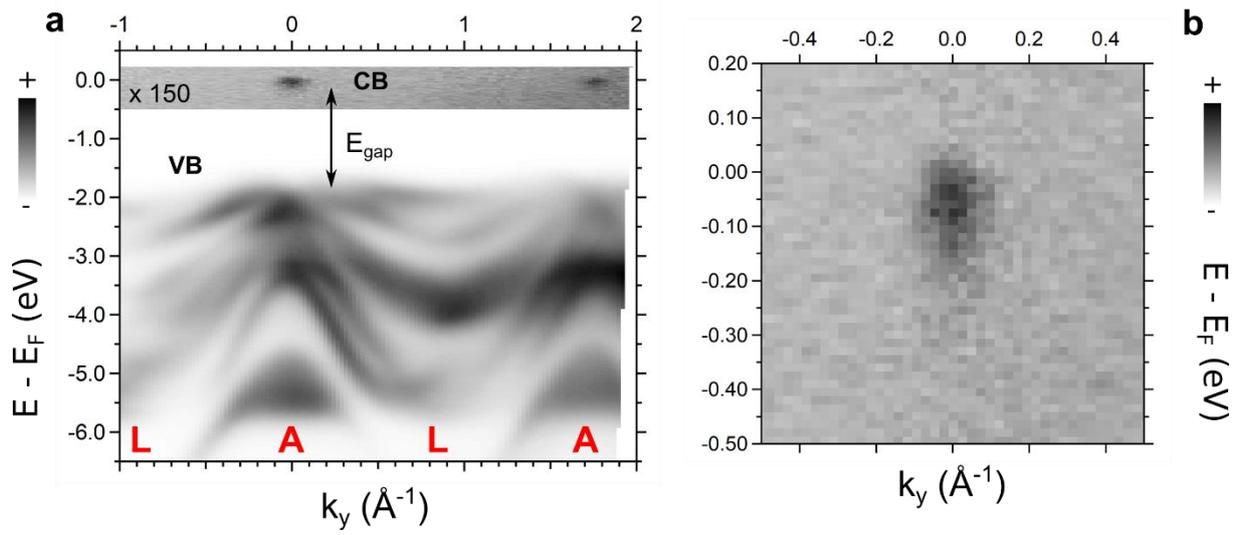

**Figure S7**: ARPES spectra of 3R α-In$_2$Se$_3$ for hν = 90 eV and T = 80 K along (a) the L-A-L high symmetry direction. It shows both the valence band and the conduction band dispersions. (b) Zoom-in close to the Fermi level.